\documentclass[letterpaper, 10 pt, conference]{ieeeconf}  

\IEEEoverridecommandlockouts                              

\usepackage{cite}

\usepackage{amsthm,amsmath,amssymb,amsfonts}
\usepackage{algorithmic}
\usepackage{graphicx}
\usepackage{textcomp}
\usepackage{xcolor}

\usepackage{sidecap}

\usepackage{bm}

\makeatletter               
\let\NAT@parse\undefined
\makeatother
\usepackage{hyperref}
\hypersetup{colorlinks,breaklinks,
linkcolor=[rgb]{0.5,0.,0.},
citecolor=[rgb]{0.000,0.427,0.173},
urlcolor=[rgb]{0.031,0.318,0.612}}

\usepackage{mathptmx}
\newtheorem{thm}{Theorem}[section]
\newtheorem{prop}[thm]{Proposition}

\usepackage{dblfloatfix}

\usepackage{booktabs} 

\def\BibTeX{{\rm B\kern-.05em{\sc i\kern-.025em b}\kern-.08em
    T\kern-.1667em\lower.7ex\hbox{E}\kern-.125emX}}

\title{\LARGE \bf
Adaptive Noise Covariance Estimation under Colored Noise using Dynamic Expectation Maximization*
}

\author{Ajith Anil Meera$^{1}$ and Pablo Lanillos$^{2}$
\thanks{*This research is supported by the Metatool project (Grant agreement 101070940) under the EIC pathfinder program. Corresponding author: ajitham1994@gmail.com
}
\thanks{$^{1,2}$ Donders Institute, Department of Artificial Intelligence, Radboud University Nijmegen, The Netherlands.}%
\thanks{$^{2}$Cajal International Center for Neuroscience, Spanish National Research Council.}
}

\begin{document}

\maketitle
\thispagestyle{empty}
\pagestyle{empty}

\begin{abstract}

The accurate estimation of the noise covariance matrix (NCM) in a dynamic system is critical for state estimation and control, as it has a major influence in their optimality. Although a large number of NCM estimation methods have been developed, most of them assume the noises to be white. However, in many real-world applications, the noises are colored (e.g., they exhibit temporal autocorrelations), resulting in suboptimal solutions. Here, we introduce a novel brain-inspired algorithm that accurately and adaptively estimates the NCM for dynamic systems subjected to colored noise. Particularly, we extend the Dynamic Expectation Maximization algorithm to perform both online noise covariance and state estimation by optimizing the free energy objective. We mathematically prove that our NCM estimator converges to the global optimum of this free energy objective. Using randomized numerical simulations, we show that our estimator outperforms nine baseline methods with minimal noise covariance estimation error under colored noise conditions. Notably, we show that our method outperforms the best baseline (Variational Bayes) in joint noise and state estimation for high colored noise. We foresee that the accuracy and the adaptive nature of our estimator make it suitable for online estimation in real-world applications.

\end{abstract}

\section{INTRODUCTION}
Identifying the noise associated with a process, i.e., estimating the Noise Covariance Matrix (NCM) is crucial for state estimation and control of a dynamic system \cite{dunik2017noise}. An incorrect NCM results in suboptimal gains (e.g., Kalman gain), significantly decreasing the quality of state estimation and tracking. Hence, accurate NCM estimation has a wide scope of applications that include robotics, signal processing, fault detection, optimal controller design, system identification, etc. However, most of the NCM estimation algorithms assume a white noise condition, which may not be true in practice. In many real-world applications the noise is colored (e.g., there are temporal autocorrelations). This makes NCM estimation under colored noise a relevant challenge \cite{liqiang2015colored}. Here, we ground on the recent advances in computational neuroscience and present a novel adaptive NCM estimator that can handle colored noise and outperforms nine state-of-the-art estimators. 

To this end, we inspire our method on the recently proposed unified theory of the brain called Free Energy Principle (FEP) \cite{friston2010free}. This mathematical foundation is particularly good at dealing with colored noise \cite{meera2020free}, as it attempts to models brain's cognitive functions under high noise temporal correlation. 
This has already attracted a large number of novel robot solutions based on the FEP---see \cite{lanillos2021active} for a review---, focused on estimation \cite{lanillos2018adaptive} and control \cite{oliver2021empirical}. 
In this work, we extend the FEP's variational inference approach for parameter estimation, named Dynamic Expectation Maximization (DEM) \cite{friston2008variational}, to perform joint online NCM and state estimation under colored noise, and show that it is highly competitive when compared to the existing estimators. 
\subsection{Contributions}
\noindent The core contributions of this paper include:
\begin{enumerate}
    \item the introduction of an online NCM estimator for linear state space systems with colored noise that converges to the global optimum of the free energy objective;
    \item the extensive evaluation of the algorithm in simulation, and benchmarking it against nine NCM estimators. 
\end{enumerate}

\subsection{Related work}
\label{sec:related}
A wide variety of NCM estimation methods have been proposed within the control community \cite{dunik2017noise}. These methods can be classified into two categories: $i)$ feedback free methods where the estimation is done by processing the entire data sequence offline and $ii)$ feedback methods where estimation is done online and. The feedback free methods are of two types: $i)$ the correlation methods that are based on the analysis of the measurement error sequence, such as  Indirect Correlation (ICM) \cite{mehra1970identification}, Input-Output Correlation (IOCM) \cite{kashyap1970maximum}, Weighted Correlation (WCM) \cite{belanger1974estimation}, Measurement Average Correlation (MACM) \cite{zhou1995estimation}, Direct Correlation (DCM) \cite{odelson2006new} and Measurement Difference Correlation (MDCM) \cite{dunik2016measurement}, and $ii)$ the Maximum-Likelihood Methods (MLM) \cite{shumway2000time} that maximises the likelihood function over the data. 

The feedback methods include Bayesian approaches, such as the Variational Bayes Method (VBM) \cite{sarkka2009recursive} and  the Covariance Matching Methods (CMM) \cite{myers1976adaptive}. However, most of these methods adhere to the white noise assumption, limiting its practical use where noises can be temporally correlated. To deal with this problem, the authors of \cite{liqiang2015colored} proposed to solve the linear equations formed from the autocovariance of the innovations of the colored noise. 
Alternatively, in this paper, we take a neuroscience-inspired approach for NCM estimation grounded on the the FEP. The closest work to our method is \cite{baioumy2022towards}, which also relies on the white noise assumption and fails to beat the benchmarks. Another approach uses an FEP based message passing scheme for noise estimation in a nonlinear system, but under white noise \cite{podusenko2022message}. The advantage of our approach when compared to the existing approaches is that it can handle noncausal colored noises. Particularly, we extend and reformulate the offline DEM algorithm, to introduce an online NCM estimator for linear systems with colored noise. We provide a quantitative analysis on the performance of our estimator and show that it outperforms nine NCM estimators under colored noise with minimal NCM estimation error. Our proposed online DEM algorithm uses free energy for online estimation, whereas the original DEM recursively iterates over the full data using free energy action as the objective for estimation \cite{friston2008variational}.

\section{PROBLEM STATEMENT} \label{sec:prob_stat}
\noindent
Consider a state space system of the form:
\begin{equation} \label{eqn:LTI_plant}
\begin{split}
        \dot{\mathbf{x}} & = \mathbf{Ax} + \mathbf{Bu} + \mathbf{w} \\
        \mathbf{y} & = \mathbf{Cx} + \mathbf{z},
\end{split}
\end{equation}
where $\mathbf{x}\in \mathbb{R}^n$ is the internal state, $\mathbf{u} \in \mathbb{R}^r$ is the control action, $\mathbf{y} \in \mathbb{R}^m$ is the output, $\mathbf{w}$ is the process noise with covariance $\mathbf{Q}$, $\mathbf{z}$ is the measurement noise with covariance $\mathbf{R}$, and $\mathbf{A}$, $\mathbf{B}$ and $\mathbf{C}$ are the matrices defining the system. The components of the plant and its estimates are denoted with and without boldface respectively. Here, $\textbf{w}$ and $\textbf{z}$ are non causal colored noises that are generated by convoluting the white noise with a Gaussain filter of kernel width $s$. The goal of this paper is to jointly estimate $R$ and $x$, given $\mathbf{y}$, $\mathbf{u}$, $\mathbf{A}$, $\mathbf{B}$, $\mathbf{C}$ and $\mathbf{Q}$.

\section{MATHEMATICAL BACKGROUND} 
This section lays out the necessary mathematical foundations for our estimator design: the noise modelling and the optimization scheme with the free energy objective. 

\subsection{Noise color modelling}
The colored noise in the plant---synthetically generated by convoluting the white noise with a Gaussian filter---presents temporal correlations, which can be captured by modelling the covariance of the noise derivatives. DEM provides a framework to seamlessly track these higher order noise derivatives using the ideas of generalized coordinates and generalized NCM. This section aims to introduce these concepts.

\subsubsection{Generalized coordinates}
The key component that enables DEM to gracefully handle colored noise is the use of generalized coordinates\footnote{Not to be confused with the definition from multibody dynamics.}. It models the evolution of the trajectory of states, instead of their point estimates. For example, the generalized states, denoted by $\tilde{x}$ is the vector collection of the higher order derivatives of $x$, given by $\Tilde{x} = [x \ x' \ x'' ...]^T$. The use of generalized coordinates provide additional information to DEM, and has proven advantageous for state estimation \cite{meera2020free} and system identification \cite{anilmeera2021DEM_LTI} under colored noise, when compared to the classical estimators, both in simulation and in real robotics problems \cite{bas2022free}. The temporal correlations in noise (because it is colored) enables the differentiation of the plant model as:
\begin{equation} 
    \begin{split}
        &x'=Ax+Bu+w \\
        &x''=Ax'+Bu'+w'\\ &...
    \end{split}
    \quad \quad
     \begin{split}
        &y=Cx+z \\
        & y' = Cx'+z'\\ &...
    \end{split}   
\end{equation}{} 
which takes the compact form: \begin{equation}
\label{eqn:generative_process}
    \begin{split}
        &\dot{\tilde{{x}}}  = D^x\tilde{{x}}= \tilde{A}\tilde{{x}}+\tilde{B}\tilde{{u}} +\tilde{{w}} 
    \end{split}{}
    \quad \quad
    \begin{split}
        &\tilde{{y}} = \tilde{C}\tilde{{x}}+\tilde{{z}}
    \end{split}{}
\end{equation}{} where $D^x = \Bigg[ \begin{smallmatrix}{}
0 & 1 & & &\\
 & 0 & 1 & & \\
 & & .& . &  \\
 & & & 0& 1 \\
 & & & & 0
\end{smallmatrix} \Bigg]_{(p+1)\times (p+1)} \otimes I_{n\times n}.$\\ 
Here, $D^x$ represents the shift matrix that performs the derivative operation on $\Tilde{x}$. $p$ and $d$ represents the embedding order (number of generalized coordinates used) of states and inputs respectively. $ \tilde A = I_{p+1} \otimes A,\hspace{10pt} \tilde B = I_{p+1} \otimes B,\hspace{10pt}\tilde C = I_{p+1} \otimes C$ represents the generalized system matrices, where $I$ denotes the identity matrix and $\otimes$ the Kronecker tensor product.

\subsubsection{Generalized noise precision}
The next key concept to handle noise color is the modelling of generalized noise precision (inverse noise covariance) matrix, defined by $\tilde{\Pi}=\Tilde{\Sigma}^{-1}$. The colored noise is assumed to be generated by the convolution of white noise by a Gaussian filter of kernel width $s$. The generalized covariance matrix for measurement noise $\Tilde{\Pi}^z = (\Tilde{\Sigma}^z)^{-1} = \mathop{\mathbb{E}}[\tilde{z}.\tilde{z}^T]$ can be derived as $\Tilde{\Pi}^z = S \otimes \Pi^z $, where $\Pi^z=R^{-1}$, and $S$ is the smoothness matrix parametrized by $s$:
\begin{equation}
\label{eqn:Smatrix}
    S(s) = \begin{bmatrix}
    1 & 0 & -\frac{1}{2s^2} & .. \\
    0 & \frac{1}{2s^2} & 0 & ..\\
    -\frac{1}{2s^2} & 0 & \frac{3}{4s^4}& ..\\
    .. & .. & .. & ..
    \end{bmatrix}^{-1}_{(p+1)\times (p+1)}.
\end{equation}
Here, $s$ denotes the noise smoothness level (kernel width of the Gaussian filter), which can be estimated online \cite{meera2022free}. $s \approx 0$ represents white noise, and high $s$ represents high noise smoothness (or color). The combined generalized noise precision $\Tilde{\Pi}$ is defined as $\Tilde{\Pi} = diag(S \otimes \Pi^z, S \otimes \Pi^w)$, where $diag(.)$ represents the block diagonal operation. The superscript notation is used to refer to the quantity under consideration and subscript notation is used to represent the derivative operation. 

\subsubsection{Parametrization of noise precision}
To facilitate an easier precision learning, we assume a zero noise cross correlation for $R$ and parameterize $\Pi^z = R^{-1}$ with a diagonal form using a hyper-parameter $\lambda = [\lambda^1 \ \lambda^2 \ ...]^T$ as:  
\begin{equation} \label{eqn:lambda_parametrization}
    \Pi^z = \begin{bmatrix}
       e^{\lambda^1} & 0 & . & .\\
       0 & e^{\lambda^2} & 0 & . \\
       0 & 0 & e^{\lambda^3} & . \\
       . & . & . & . & .
    \end{bmatrix}
\end{equation}
The exponential parametrization was chosen to ensure that the precision is always positive definite, $\Pi^z \succ 0$. 

\subsection{Free energy optimization}
The estimation is formulated as an optimization problem where the agent minimizes its free energy objective, an information theoretic quantity that bounds surprise~\cite{friston2008variational}. This section aims to derive the mathematical form of this objective to estimate the NCM and the states as defined in Section \ref{sec:prob_stat}.


\subsubsection{Estimation as Bayesian inference}
The core tenet of FEP is that it describes the agent as a Bayesian machine that  estimates the posterior probability of the parameters $\vartheta$ using the measurements $\textbf{y}$. Using Bayes rule, the posterior is $p(\vartheta/\textbf{y}) = {p(\vartheta,\textbf{y}})/{\int p(\vartheta,\textbf{y})d\vartheta}$. Since the denominator contains an intractable integral, variational methods use a recognition density $q(\vartheta)$ that closely approximates the posterior as $q(\vartheta)\approx p(\vartheta/\textbf{y})$. This involves minimizing the Kullback-Leibler (KL) divergence of the distributions,  $KL(q(\vartheta)||p(\vartheta/\textbf{y})) = \langle \ln q(\vartheta)\rangle_{q(\vartheta)} - \langle \ln{p(\vartheta/\textbf{y})}\rangle_{q(\vartheta)}$, where $\langle.\rangle_{q(\vartheta)}$ represents the expectation over $q(\vartheta)$. Rearranging the terms and using $p(\vartheta/\textbf{y}) = {p(\vartheta,\textbf{y})}/{p(\textbf{y})}$ yields \cite{friston2010free}: 
\begin{equation}    
    \label{eqs:logevidence}
    \begin{split}
     \ln  p(y) & =  F + KL(q(\vartheta)||p(\vartheta|\textbf{y})), \\
        F = & \langle \ln{p(\vartheta,\textbf{y})}\rangle_{q(\vartheta)}  -\langle \ln q(\vartheta)\rangle_{q(\vartheta)}
    \end{split}
\end{equation}
where $F$ is called the free energy. From Equation \ref{eqs:logevidence}, the minimization of KL divergence results in the maximization of free energy, as $\ln p(\textbf{y})$ is independent of $\vartheta$. This is the core idea behind using free energy optimization as a proxy for brain's inference, thereby minimizing the brain's sensory surprisal \cite{friston2010free}.

\subsubsection{Free energy objective}
To compute the free energy objective,
we make two fundamental assumptions about the recognition density $q(\vartheta) = q(\tilde{x},\lambda)$: i) Mean field assumption \cite{friston2008variational} that facilitates a conditional independence between the subdensities, $q(\vartheta) = q(\tilde{x})q(\lambda)$,  and ii) Laplace assumption \cite{friston2007variational} that facilitates the use of Gaussian distributions with mean $\mu$ and variance $\Sigma$ over these subdensities, $q(\tilde{x})= \mathcal{N}(\tilde{x}:\mu^{\tilde{x}},\,\Sigma^{\tilde{x}})$ and $q(\lambda)= \mathcal{N}(\lambda:\mu^{\lambda},\,\Sigma^{\lambda})$. Laplace assumption is made on the generative model of the plant, which generates the data as $p(\Tilde{\mathbf{y}}/\vartheta) = \mathcal{N}(\tilde{y}: \tilde{C} \tilde{x},R), \ p(\tilde{x}'/ 
\lambda) =  \mathcal{N}(D^x \Tilde{x}: \tilde{A}\tilde{x}+\tilde{B}\tilde{u} ,Q)$, and the prior distribution of $\lambda$ as $p(\lambda) =  \mathcal{N}(\lambda:\eta^\lambda,(P^\lambda)^{-1})$, where $\eta^\lambda$ is the prior noise parameter with prior precision $P^\lambda$. Under these assumptions, $F$ from Equation \ref{eqs:logevidence} upon further simplification reduces to the sum of three terms: $i)$ precision weighted prediction errors, $ii)$ information entropy, and $iii)$ mean field terms. The full free energy objective after dropping constants become:
\begin{equation} \label{eqn:F_bar_MF}
        {F} =  
                  U(y,\vartheta) + H^{\tilde{x}} + H^\lambda 
                  + W^{\Tilde{x}} + W^{\lambda}. 
\end{equation}
where $H^{\Tilde{x}}$ and $H^\lambda$ are the entropy terms given by:
\begin{equation}
    H^{\tilde{x}} = \frac{1}{2}  \ln |\Sigma^{ \tilde{x}}|, \ H^\lambda = \frac{1}{2} \ln |\Sigma^\lambda| ,
\end{equation}
$W^{\Tilde{x}}$ and $W^{\lambda}$ are the mean field terms given by:
\begin{equation}
    W^{\tilde{x}} = \frac{1}{2}     tr \big[ \Sigma^{ \tilde{x}} U(y,\mu^\vartheta)_{ \tilde{x} \tilde{x}} \big], \ W^{\lambda} = \frac{1}{2}     tr \big[ \Sigma^{ \lambda} U(y,\mu^\vartheta)_{ \lambda \lambda} \big],
\end{equation}
$U(y,\vartheta)$ is the internal energy given by:
\begin{equation} \label{eqn:U(theta,y)}
\begin{split}
        U(\vartheta,y)  = &  -\frac{1}{2} {\epsilon}^{\lambda T}P^\lambda {\epsilon}^\lambda + \frac{1}{2} \ln |P^\lambda|  - \frac{1}{2}  \tilde{\epsilon}^{T}  \tilde{\Pi}  \tilde{\epsilon} + \frac{1}{2} \ln | \tilde{\Pi}|
\end{split}
\end{equation}
and $\tilde{\epsilon}$ is the prediction error given by:
\begin{equation}
    \tilde{\epsilon} = \begin{bmatrix}
\tilde{\epsilon}^{{y}} \\  \tilde{\epsilon}^{ {x}}
\end{bmatrix} = 
\begin{bmatrix}
 \tilde{\textbf{y}} -  \Tilde{C} \tilde{x}  \\ D^x{\tilde{x}} -  \tilde{A} \Tilde{x} - \Tilde{B} \tilde{u}
\end{bmatrix}.
\end{equation}
We refer to \cite{anilmeera2021DEM_LTI,friston2008variational} for an elaborate read on similar simplifications. The free energy form in Equation \ref{eqn:F_bar_MF} will be used for the estimator design in the next section.

\section{PROPOSED ADAPTIVE ESTIMATOR}
In this section, we introduce our NCM estimator design, alongside its guarantees for a global free energy optima, to finally propose a joint state and NCM estimator.

\subsection{Noise covariance matrix (NCM) estimation} \label{sec:noise_cov_est}
According to FEP, the brain's inference process proceeds through the gradient ascend on its free energy objective $F$. Inspired by this idea, we introduce a novel adaptive NCM estimator using the first two gradients of $F$ denoted by $F_{\lambda}$ and $F_{\lambda \lambda}$. We use the Gauss-Newton update scheme \cite{ozaki1992bridge} that maximises $F$ by making incremental updates to the noise parameter estimate $\lambda$ as:
\begin{equation} \label{eqn:lambda_update}
      \lambda(t+ \Delta t)  = \lambda(t) + \big[ (e^{  F_{\lambda \lambda}  \Delta t}-I) (F_{\lambda \lambda})^{-1} F_{\lambda} \big] \Big|_{\lambda = \lambda(t)  },
\end{equation}
where $\lambda(t)$ is the noise parameter at time $t$ and $\Delta t$ is the sampling time. The gradient of $F$ can be evaluated by differentiating Equation \ref{eqn:F_bar_MF} with $\lambda$, which simplifies as:
\begin{equation} \label{eqn:F_lambda}
\begin{split}
    F_{ \lambda}  = & - \frac{1}{2} (\Tilde{\epsilon}^T  \Tilde{\Pi} \Tilde{\epsilon} )_\lambda - \epsilon^{\lambda T} P^\lambda  + p+1 \\
    & -\frac{1}{2}     tr \big(  \Sigma^{ {x}} (\tilde{\epsilon}^{T}_{\Tilde{x}} \Tilde{\Pi} \tilde{\epsilon}_{\Tilde{x}} )_{\lambda} \big)   - \frac{1}{4} tr \big( \Sigma^\lambda  (\tilde{\epsilon}^{T} \Tilde{\Pi} \tilde{\epsilon})_{\lambda \lambda \lambda} \big) 
    \end{split}.
\end{equation}
The exponential parametrization of $\Pi^z$ with respect to $\lambda^i$ in Equation \ref{eqn:lambda_parametrization} results in the same higher order gradients ($\Tilde{\Pi}_{\lambda^i } = \Tilde{\Pi}_{\lambda^i \lambda^i }= \Tilde{\Pi}_{\lambda^i \lambda^i \lambda^i}  $), which simplifies $F_{\lambda^i \lambda^i}$ as:
\begin{equation} \label{eqn:F_hh}
\begin{split}
            F_{{\lambda^i} {\lambda^i}} =  - \frac{1}{2}\Tilde{\epsilon}^T  \Tilde{\Pi}_{\lambda^i } \Tilde{\epsilon}  - P^{\lambda^i \lambda^i} &- \frac{1}{2}     tr \big(  \Sigma^{ {x}} \tilde{\epsilon}^{T}_{\Tilde{x}} \Tilde{\Pi}_{\lambda^i} \tilde{\epsilon}_{\Tilde{x}} \big) \\
            & - \frac{1}{4} tr \big( \Sigma^{\lambda}  \tilde{\epsilon}^{T} \Tilde{\Pi}_{\lambda^i} \tilde{\epsilon} \big) .
\end{split}
\end{equation}
Note that $F_{\lambda^i \lambda^j}=0 \ \forall \ i \neq j$ since $\tilde{\Pi}_{\lambda^i \lambda^j} = O \ \forall \ i \neq j$. Therefore, the Hessian matrix $F_{\lambda \lambda}$ is diagonal, aiding a faster computation. The use of free energy maximization for the estimator design is motivated by the existence of a unique maximum for the free energy curve given in Equation \ref{eqn:F_bar_MF} with the  right choice of $\eta^\lambda$ and $P^\lambda$, which guarantees the convergence of the estimator without instability.

\begin{prop}  \label{prop:convergence}
Free energy objective $F$ defined by Equation \ref{eqn:F_bar_MF} has a global maximum with respect to the noise parameter $\lambda$ (using the exponential parametrization in Equation \ref{eqn:lambda_parametrization}), for all $\bm{\lambda^i}$ of data, under the choice of prior $\eta^{\lambda}$ and a diagonal prior precision $P$ that satisfies the condition  $\bm{\lambda^i} < \eta^{\lambda^i}  + \frac{p+1}{P^{\lambda^i \lambda^i}} +1 $.
\end{prop}
\begin{proof}
There is a global maximum for $F$ with respect to $\lambda$ if we can prove that $F_{\lambda \lambda} \prec O \ \forall \ \lambda$ where $F_{\lambda}=0$. Equating Equation \ref{eqn:F_lambda} to zero yields the condition satisfied by all $\lambda$ with zero gradients ($F_{\lambda^i}=0$) as:
\begin{equation}
\begin{split}
    - \frac{1}{2} \Tilde{\epsilon}^T  \Tilde{\Pi}_{\lambda^i} \Tilde{\epsilon} =   \epsilon^{\lambda T} P^\lambda \epsilon^{\lambda}_{\lambda^i}   - p-1 + \frac{1}{2}     tr \big(  \Sigma^{ {x}} \tilde{\epsilon}^{T}_{\Tilde{x}} \Tilde{\Pi}_{\lambda^i} \tilde{\epsilon}_{\Tilde{x}}  \big)  \\
    + \frac{1}{4} tr \big( \Sigma^\lambda  \tilde{\epsilon}^{T} \Tilde{\Pi}_{\lambda^i} \tilde{\epsilon} \big) 
\end{split}
\end{equation}
Substituting it in Equation \ref{eqn:F_hh} and enforcing $F_{\lambda^i \lambda^i}<0$ gives the condition for the existence of a global maximum for $F$ with respect to $\lambda$ as $ \epsilon^{\lambda T} P^\lambda \epsilon^{\lambda}_{\lambda^i}   - p-1 <0$, which upon rearranging yields $ \lambda^{i} < \eta^{\lambda^i} + \frac{p+1}{P^{\lambda^i \lambda^i}} + 1$. Assuming that the maximum of $F$ is close to the real $\bm{\lambda^i}$ of data ($\lambda^{i} \approx \bm{\lambda^i}$, later shown in simulation), the choice of $\eta^\lambda$ and $P^\lambda$ that sufficiently ensures the existence of global maximum for $F$ with respect to $\lambda$ is given by:
\begin{equation} \label{eqn:cond_maximum}
    \bm{\lambda^i} < \eta^{\lambda^i} + \frac{p+1}{P^{\lambda^i \lambda^i}} + 1.
\end{equation} 
 For example, the choice of $\eta^\lambda = 0$ and $P^{\lambda\lambda}=1$ with $p=6$ is sufficient to ensure the existence of a global maximum for $F$ with respect to $\lambda \ \forall \ \bm{\lambda^i}<8$ of the data. This completes the proof.

\end{proof}

\subsection{Joint state and noise covariance observer design} \label{sec:joint_est}
This section aims to introduce a novel observer design for the joint state and noise covariance estimation for linear systems with colored noise. We use our previously proposed DEM based state observer given by \cite{meera2020free}:
\begin{equation} \label{eqn:DEM_state_update}
    \dot{\tilde{x}} =     A_1 \tilde{x} +  
    B_1 \begin{bmatrix}
    \tilde{\textbf{y}} \\ \tilde{\textbf{u}}
    \end{bmatrix}  
\end{equation}
where $A_1 = [D^x-k^x\tilde{C}^T\tilde{\Pi}^z\tilde{C}  -k^x(D^x-\tilde{A})^T \tilde{\Pi}^w(D^x-\tilde{A})]$, $B_1 = k^x\begin{bmatrix}
\tilde{C}^T\tilde{\Pi}^z & (D^x-\tilde{A})^T\tilde{\Pi}^w\tilde{B}
\end{bmatrix}$, and $k^x$ is the learning rate. We combine this estimator with our noise covariance estimator from the previous section to jointly estimate $\tilde{x}$ and $\lambda$. Since Equation \ref{eqn:DEM_state_update} is a linear differential equation, the exact discretization of the observer is:
\begin{equation} \label{eqn:state_update_rule}
    \tilde{x}(t+dt) = e^{A_1 dt} \tilde{x}(t) + A_1^{-1}(e^{A_1 dt} - I)B_1 \begin{bmatrix}
    \tilde{\textbf{y}}(t) \\ \tilde{\textbf{u}}(t)
    \end{bmatrix}.  
\end{equation}
Equations \ref{eqn:lambda_update} and \ref{eqn:state_update_rule} together completes the joint observer design. Since the update rule in Equation \ref{eqn:lambda_update} is nonlinear in $\lambda$, the stability proof for the joint estimator is non trivial, and is left for future work. Importantly, there is a closed-form solution for the conditional covariance of the joint estimator (given in Proposition \ref{prop:cond_cov}) that facilitates the potential use of our estimator for uncertainty based decision making and planning.

\begin{prop}  \label{prop:cond_cov}
The free energy optimization based joint state and noise covariance estimation scheme introduced in Section \ref{sec:joint_est} has a conditional covariance for $\lambda$ and $\tilde{x}$ given by $\Sigma^{\lambda} = \mathop{\mathbb{E}}[\lambda. \lambda^T]  = (P^{\lambda} + \frac{1}{2} (\epsilon^T \Pi \epsilon)_{\lambda \lambda} )^{-1}$ and $\Sigma^{\tilde{x}} = \tilde{\epsilon}^{T}_{\Tilde{x}} \Tilde{\Pi} \tilde{\epsilon}_{\Tilde{x}}$ respectively. 
\end{prop}
\begin{proof}
The free energy optimization based estimator design tracks the maximum of free energy for estimation. The gradient of $F$ along the conditional covariance of $\lambda$ is zero at the peaks of $F$, $F_{ \Sigma^\lambda} = 0$. Differentiating Equation \ref{eqn:F_bar_MF} with respect to $\Sigma^\lambda$ gives:
\begin{equation}
    \frac{\partial F}{\partial \Sigma^\lambda} = \frac{1}{2} \frac{\partial}{\partial \Sigma^\lambda}(ln|\Sigma^\lambda| + tr(\Sigma^\lambda U(y,\mu^\vartheta)_{\lambda \lambda})),
\end{equation}
which when equated to zero yields the optimal conditional precision as:
\begin{equation}
    (\Sigma^\lambda )^{-1} = -U(y,\mu^\vartheta)_{\lambda \lambda} = P^\lambda + \frac{1}{2} (\epsilon^T \Pi \epsilon)_{\lambda \lambda}.
\end{equation}
Similarly, $\Sigma^{\tilde{x}}$ can be evaluated by setting $F_{\Sigma^{\tilde{x}}}=0$, giving the optimal conditional covariance of $\tilde{x}$ as $(\Sigma^{\tilde{x}})^{-1} = - U(y,\mu^\vartheta)_{\tilde{x}\tilde{x}}$, which upon simplification using Equation \ref{eqn:U(theta,y)} yields:
\begin{equation}
    \Sigma^{\tilde{x}} =  ( \tilde{\epsilon}^{T}_{\Tilde{x}} \Tilde{\Pi} \tilde{\epsilon}_{\Tilde{x}} )^{-1}.
\end{equation}
Note that the mean field assumption that separated the recognition densities $q(\vartheta) = q(\tilde{x})q(\lambda)$, directly results in $\Sigma^{\tilde{x}\lambda} = O$. This completes the proof.
\end{proof}


\section{BENCHMARKING}
Using randomized simulation experiments, we compare our proposed online DEM against nine state-of-the-art estimators (See Sec.~\ref{sec:related}): VBM, ICM, IOCM, WCM, MACM, MLM, DCM, MDCM and CMM.

\subsection{Numerical simulations settings} \label{sec:sim_setting}
The following simulation settings are used throughout the paper, unless mentioned otherwise. The system in Equation \ref{eqn:LTI_plant} with a randomly sampled stable $A$ matrix with $n=2$ is used. $B = O_{2 \times 1}$ and $C = I_{2 \times 2}$ with noise covariance $Q=e^{-5}I_{2\times 2}$, $R=diag(e^{-4}, e^{-3})$. This particular system was chosen so that all the benchmarks could be used for the comparison. The data was generated with $dt=0.1s$, initial state $x_0 = [ \begin{smallmatrix}
    1 \\ -1
\end{smallmatrix} ]$, $s=0.5$ until $T=32s$. An uninformative prior of $\eta^{\lambda^i}=0.001$ with $P^{\lambda} = e^{-1}I_{2 \times 2}$ was placed on the noise parameter $\lambda^i$. The embedding order of $p=6$ and $d=2$ were used for data generation and the estimator. The MATLAB code from \cite{dunik2017noise} was adapted to simulate the benchmarks. The corresponding simulation settings are given in Appendix \ref{sec:sim_benchmarks}. Code available at https://github.com/ajitham123/DEM\_NCM.



\subsection{Noise covariance estimator}
We benchmarked our NCM estimator for white noise and colored noise in highly randomized experimental simulations.

\begin{figure}[!hbtp]
\centering
\includegraphics[scale = 0.32]{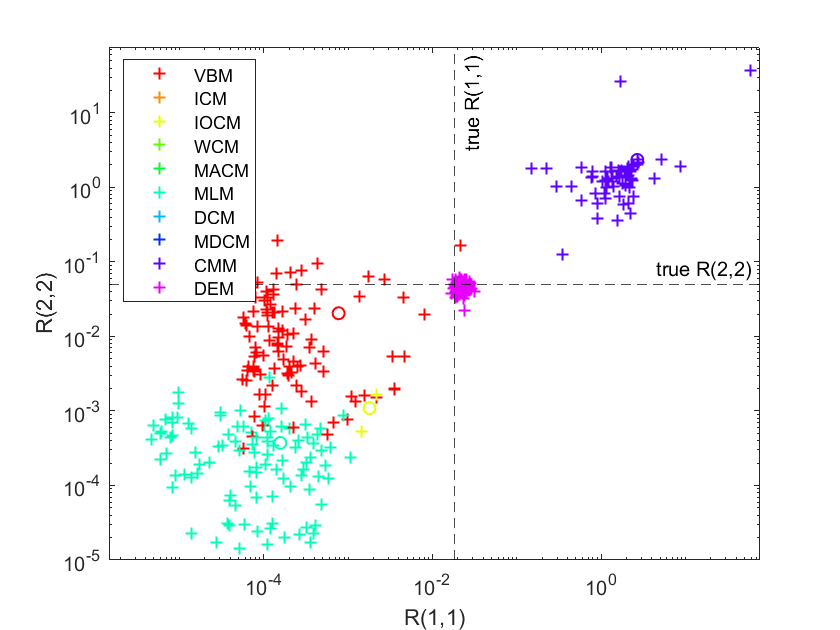}
\caption{Noise covariance estimation under colored noise. While most methods provide unstable solutions, our online DEM consistently produces a stable solution (in pink), which outperforms all other stable benchmarks under colored noise - VBM (in red), CMM (in blue) and MLM (in cyan).}
\label{fig:bench_hz_color}
\end{figure}

Figure \ref{fig:bench_hz_color} shows the results of 100 randomized trials with colored noise ($s=0.5$) case with the same setup as that of Section \ref{sec:sim_setting}. While most of the benchmarks provided an unstable (exploding) solution except for VBM (in red), MLM (in cyan) and CMM (in blue), online DEM (in pink) provided a consistently stable solution with high accuracy. 

\begin{figure}[!hbtp]
\centering
\includegraphics[scale = 0.3]{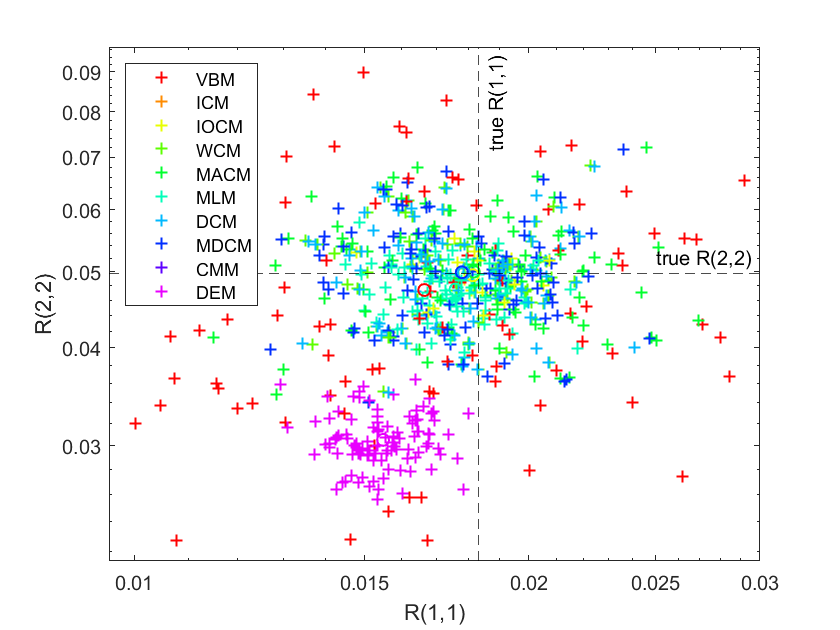}
\caption{The covariance $R$, estimated by the benchmarks for 100 randomized trials with white noise. The $A$ matrix was randomly selected. While most benchmarks provide an unbiased estimate, online DEM (in pink) provides a biased estimate (because of priors), but with acceptable levels of accuracy. }
\label{fig:bench_hz_white}
\end{figure}

Figure \ref{fig:bench_hz_white} shows the simulation results of 100 randomized trials with the setup given in Section \ref{sec:sim_setting} for the white noise case with $s=10^{-10}$ and $P^\lambda = e^{0}I_{2 \times 2}$. While most benchmarks provide an unbiased estimate of R (true value at the centre of the dashed cross), online DEM (in pink) provides a biased estimate with acceptable levels of accuracy. 

\begin{table}[htbp]
\caption{Error in noise covariance matrix estimation (SSE)}
\begin{center}
\begin{tabular}{ccc}
\toprule
 & \textbf{White noise} & \textbf{Colored noise} \\
\midrule
\textbf{VBM}  & (27.42 $\pm$ 31.44) $\times 10^{-5}$ &  (2.303 $\pm$ 2.724) $\times 10^{-3}$ \\ 
\textbf{ICM}  & (5 $\pm$ 6.14) $\times 10^{-5}$ &  Inf \\ 
\textbf{IOCM} & (1.25 $\pm$ 1.07) $\times 10^{-5}$  &  (2.644 $\pm$ 0.095) $\times 10^{-3}$ \\
\textbf{WCM}  & (5 $\pm$ 6.14) $\times 10^{-5}$ &  Inf  \\
\textbf{MACM} & (7.03 $\pm$ 8.3) $\times 10^{-5}$  &  Inf \\
\textbf{MLM} &   (2.12 $\pm$ 2.47) $\times 10^{-5}$  &  (2.772 $\pm$ 0.039) $\times 10^{-3}$   \\
\textbf{DCM}  &  (4.99 $\pm$ 6.14) $\times 10^{-5}$ &  Inf \\
\textbf{MDCM} &  (6.16 $\pm$ 8.05) $\times 10^{-5}$ &  Inf   \\
\textbf{CMM} &  62.01 $\pm$ 140.7  &  6.501 $\pm$ 10.81 \\
\textbf{DEM}  & (38.65 $\pm$ 9.11) $\times 10^{-5}$  &  (9.344 $ \pm$ 9.892) $\times 10^{-5}$ \\
\bottomrule
\end{tabular}
\label{tab:Error_est}
\end{center}
\end{table}

Table \ref{tab:Error_est} reports the average error in NCM estimation (sum of squared error) for all the benchmarks, for both white noise and colored noise case. Our proposed online DEM outperforms other methods by 2 orders of magnitude for the colored noise case with minimal estimation error. Table \ref{tab:unstable_soln} reports the percentage of stable solutions provided by all benchmarks, highlighting the result that DEM consistently provides a stable solution. In summary, DEM outperforms the benchmarks in NCM estimation under colored noise.

\begin{figure}[!htb]
\centering
\includegraphics[scale = 0.31]{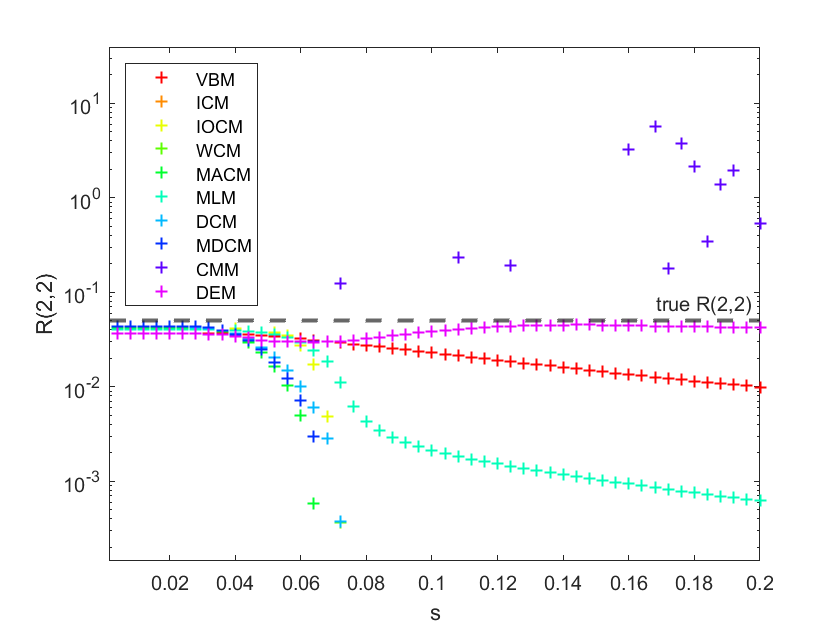}
\caption{The influence of noise color on the stability of benchmarks. With an increasing noise smoothness $s$, all methods become unstable (after around $s=0.06$), except for DEM, VBM, CMM and MLM. DEM (in pink) remains closest to the true $R$ value (in dashed black). }
\label{fig:sensitivity_to_s}
\end{figure}

\begin{table}[!htb]
\caption{Percentage of stable solutions }
\begin{center}
\begin{tabular}{ccc}
\toprule
 & \textbf{White noise} & \textbf{Colored noise} \\
\midrule
\textbf{VBM}  & 100\% & 100\% \\ 
\textbf{ICM}  & 100\% & 0\% \\ 
\textbf{IOCM} & 19\% & 2\% \\
\textbf{WCM}  & 100\% & 0\% \\
\textbf{MACM} & 100\% & 0\% \\
\textbf{MLM} &  100\% & 100\%  \\
\textbf{DCM}  & 100\% & 0\% \\
\textbf{MDCM} &  100\% & 0\%   \\
\textbf{CMM} &  12 \% & 59\% \\
\textbf{DEM}  & 100\% &  100\% \\
\bottomrule
\end{tabular}
\label{tab:unstable_soln}
\end{center}
\end{table}

\subsection{The influence of noise color on estimation}
This section aims to analyze the performance of DEM with increasing levels of noise color and, to compare it against the benchmarks. The same simulation setup from Section \ref{sec:sim_setting} with a randomly sampled matrix, $A = [ \begin{smallmatrix}
    0.0484 & 0.7535 \\ -0.7617  & -0.2187
\end{smallmatrix} ]$ was used for increasing levels of noise smoothness $s$ from 0 to 0.2, with $P^\lambda = e^0 I_{2 \times 2}$. Figure \ref{fig:sensitivity_to_s} shows the high sensitivity of all benchmarks with increasing $s$ (around $s=0.06$), while DEM (in pink) shows consistent performance by remaining close to the true $R$ (in dashed black). This concludes that DEM can seamlessly handle noise color, while most of the benchmark demonstrates a high sensitivity.

\begin{figure}[!h]
\centering
\includegraphics[scale = 0.3]{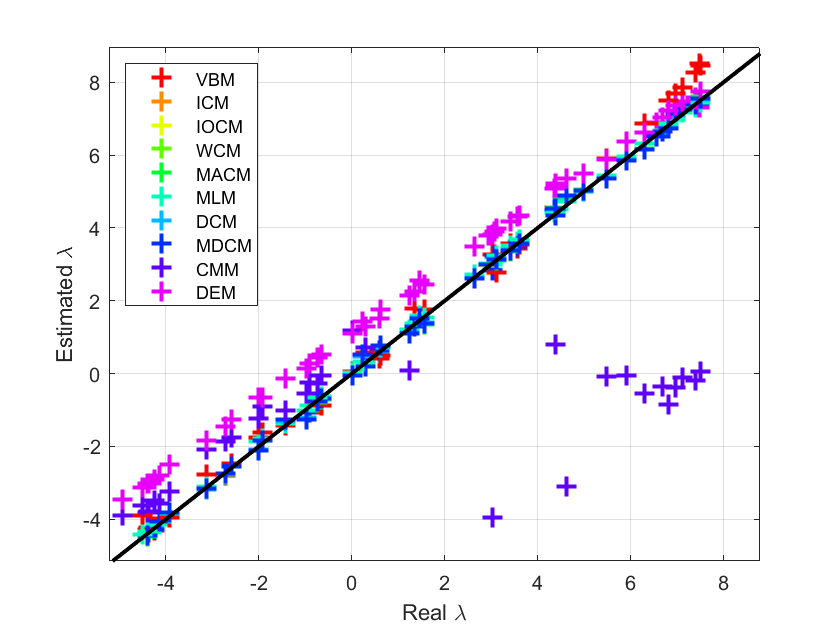}
\caption{The performance of estimators for different noise levels $\lambda$, under white noise condition. All methods except CMM provides a solution that is close to the real $\lambda$ (colored points are close to the black line). DEM provides a biased solution.   }
\label{fig:lambda_random_white}
\end{figure}

\begin{figure}[!h]
\centering
\includegraphics[scale = 0.29]{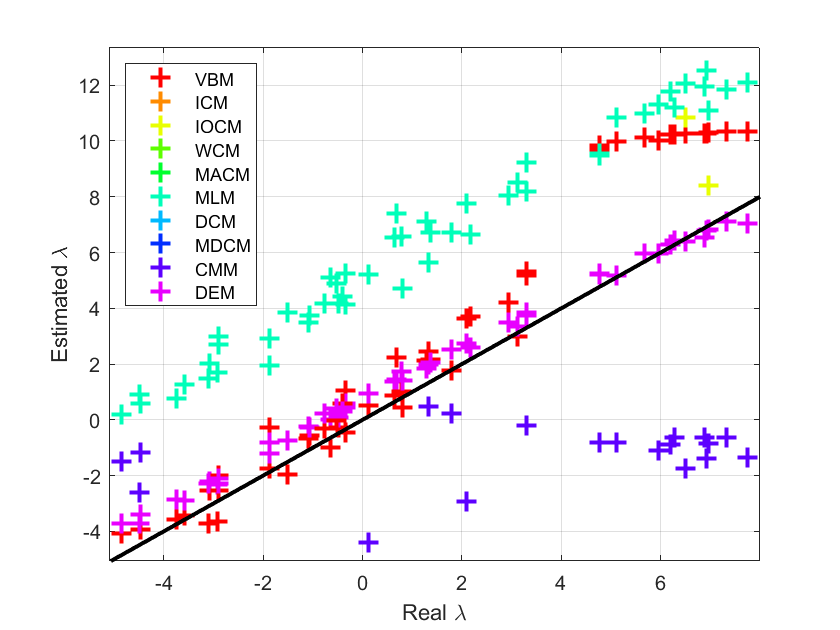}
\caption{Under colored noise ($s=0.5$), DEM outperforms all benchmarks with an accurate estimate of $\lambda$ (pink points are the closest to the black line) for all ranges of $\lambda$. Under colored noise, VBM (in red) is DEM's main competitor.}
\label{fig:lambda_random}
\end{figure}

\subsection{The influence of noise levels on estimation}
The previous sections focused on demonstrating the effectiveness of online DEM in estimating noise CM for a fixed $R$. This section aims to analyse the performance of DEM for varying $R$. 50 randomised trials with the setup from Section \ref{sec:sim_setting} was used by varying $\lambda$ from -5 to 8, and the results are shown in Figure \ref{fig:lambda_random_white} and \ref{fig:lambda_random} for white noise and colored noise respectively. The best method is expected to remain as close as possible to the black line. Online DEM remains close to the black line for varying $R$, both for white noise and colored noise, while most benchmarks show an inaccurate estimation for colored noise case. It can be concluded from Figure \ref{fig:lambda_random} that the best competitor for online DEM for colored noise case is VBM (in red).

\begin{figure}[!h]
\centering
\includegraphics[scale = 0.27]{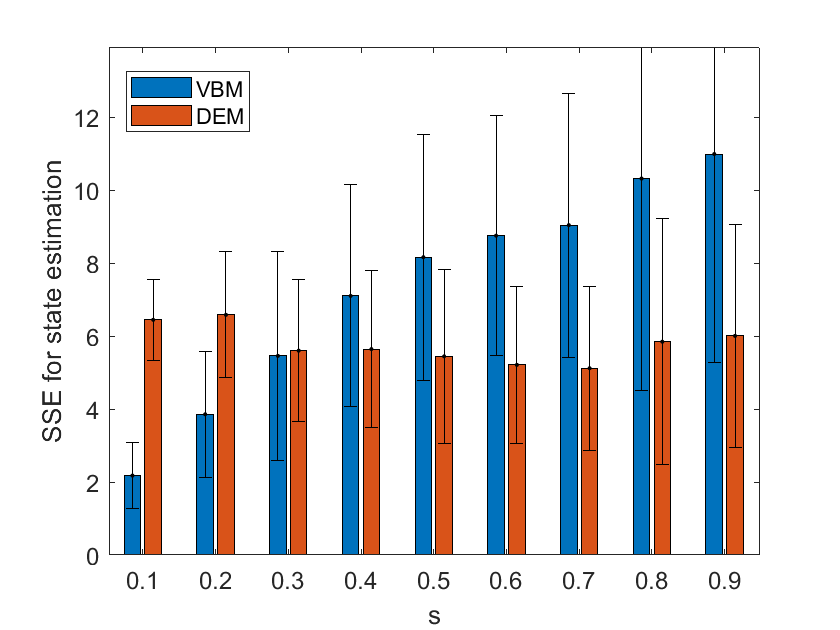}
\caption{The SSE for state estimation for 100 randomized trials each for different noise smoothness $s$. Online DEM (in red) outperforms VBM (in blue) for high colored noise ($s>0.3$) with a lower SSE for state estimation.}
\label{fig:SSEx_vs_smoothness}
\end{figure}

\subsection{State observer}
\begin{figure}[!h]
\centering
\includegraphics[scale = 0.27]{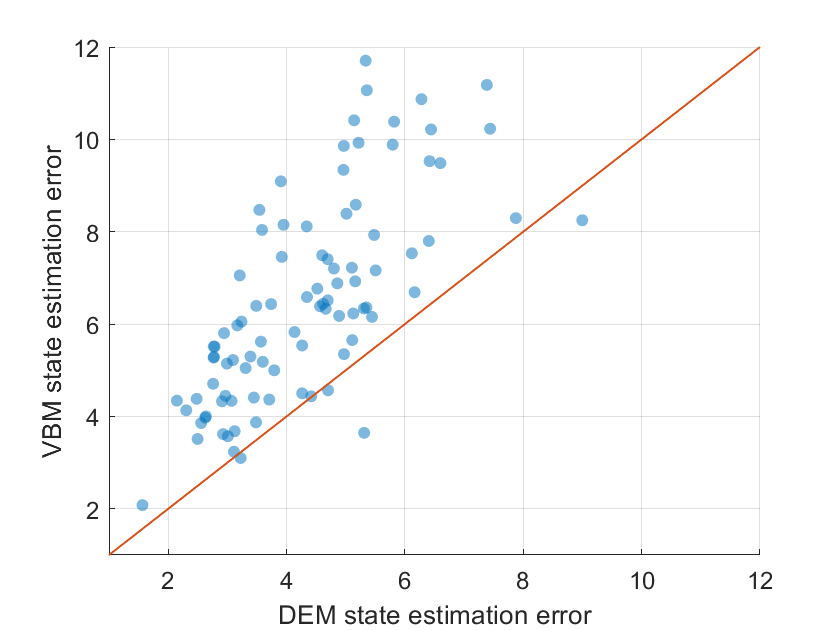}
\caption{The comparison of joint state and NCM estimation of DEM and VBM. Since most of the points lie above the red line, DEM is highly reliable for online state estimation under high colored noise. }
\label{fig:VBM_DEM_state_100}
\end{figure}

This section aims to analyse the performance of our joint state and noise CM estimator for colored noise, in comparison with the best competing benchmark (VBM). Figure \ref{fig:SSEx_vs_smoothness} shows the SSE for state estimation under varying noise color for online DEM and VBM, under 100 randomized trials as per Section \ref{sec:sim_setting} with varying $s$ from 0.1 to 0.9. For higher noise color ($s>0.3$), online DEM outperforms VBM with minimal error in state estimation. For lower noise color, the state estimation of DEM is influenced by the higher noise derivatives and underperforms VBM. Figure \ref{fig:VBM_DEM_state_100} shows the results of 100 randomized trials with $s=0.5$ from Figure \ref{fig:SSEx_vs_smoothness}. Since most points lie above the red line, DEM is highly reliable for higher noise color when compared to VBM. In summary, our proposed online DEM beats the best benchmark for the state estimation under higher noise color.

\subsection{Transient noise covariance}
\begin{figure}[htpb]
\centering
\includegraphics[scale = 0.27]{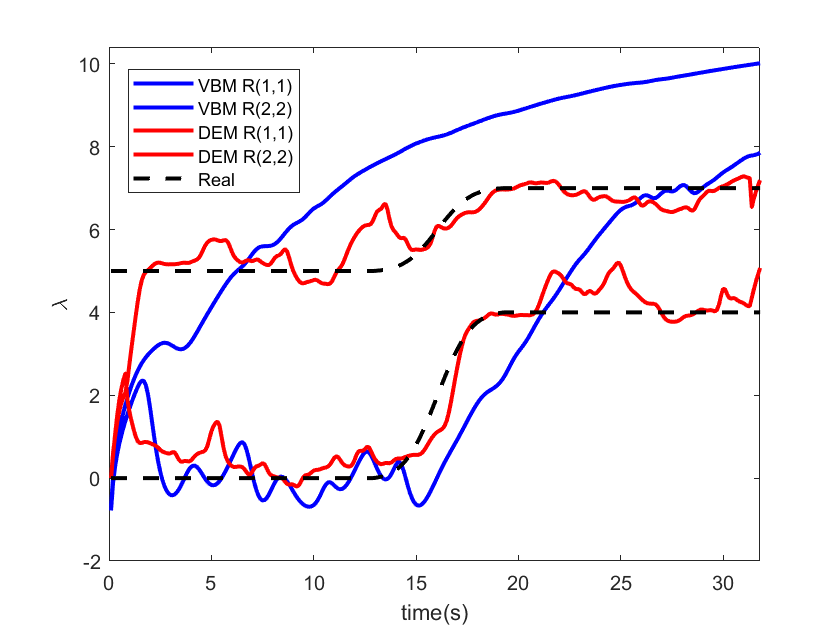}
\caption{The quality of online estimation of $\lambda$. DEM (in red) closely approximates the real $\lambda$ (in black), when compared to VBM (in blue).   }
\label{fig:online_noise_est}
\end{figure}

This section aims to analyze the performance of DEM and VBM under transient noise CM. The setup from Section \ref{sec:sim_setting} with  $A = [ \begin{smallmatrix}
    0.0484 & 0.7535 \\ -0.7617  & -0.2187
\end{smallmatrix} ]$ was used with a varying $\lambda$, and the results are plotted in Figure \ref{fig:online_noise_est}. It can be observed that DEM (in red) tracks the true $\lambda$ (in dotted black) with respect to time, when compared to VBM (in blue). In summary, DEM outperforms VBM for online noise CM estimation under colored noise condition.

\begin{figure}[h]
\centering
\includegraphics[scale = 0.26]{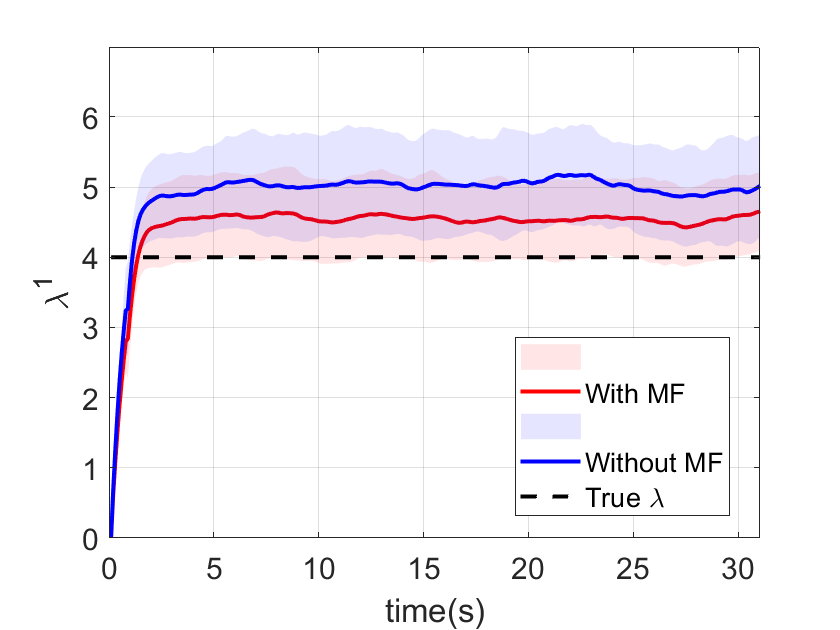}
\caption{The influence of mean field terms on the noise estimation. The average of 100 randomised trials with (in red) and without (in blue) using mean field terms is plotted. Using the mean field terms improves the noise estimation (red curve is closer to the black curve than the blue curve).  }
\label{fig:DEM_MF_terms}
\end{figure}

\subsection{Influence of the mean field terms on estimator}
The mean field terms ($W$ in Equation \ref{eqn:F_bar_MF}) are often neglected in practice during estimation and control \cite{lanillos2021active}. This section aims to draw attention to the influence of such an approximation on the quality of NCM estimation. Figure \ref{fig:DEM_MF_terms} shows the results of 100 randomized trials for online NCM estimation, with (in red) and without (in blue) using the mean field terms during estimation. The red curve is closer to the black curve than the blue curve, indicating that the use of mean field term improves the quality of NCM estimation.

\section{CONCLUSION}
A novel and adaptive joint state and NCM estimator (online DEM) for linear systems with colored noise was introduced, grounding on the neuroscientific theory of FEP. Our NCM estimator was proven to converge to the global optimum of the free energy objective. Under randomized numerical simulations, online DEM was shown to outperform nine NCM estimators with minimal estimation error under colored noise. While most estimators displayed instability under colored noise, online DEM provided a consistently superior performance. The main drawback of our method is its higher computational complexity due to the use of generalized coordinates. Future work will focus on applying the online DEM on robotic system with transient noise characteristics (e.g., delivery drone flying in wind). Furthermore, the mathematical framework introduced in this paper can be extended to estimate $Q$ online.

\addtolength{\textheight}{-12cm}   



\section*{APPENDIX}

\subsection{Simulation setup for benchmarks} \label{sec:sim_benchmarks}
This section provides the simulation conditions used by the benchmarks. The constants have the same definition as in \cite{dunik2017noise}. All the benchmarks were provided with the correct initial condition of $x_0 = [ \begin{smallmatrix}
    1 \\ -1
\end{smallmatrix} ]$. A stable filter gain of $K = 0.8 I_{2 \times 2}$ was used for ICM, WCM and DCM. The weight matrices for MACM were chosen as $W_1 = I_{4 \times 4}$ and $W_2 = I_{6 \times 6}$. MLM used a maximum number of EM steps of 50, with an initial $R_a = [ \begin{smallmatrix}
    2.47 & -0.58 \\ -0.58 & 3.37
\end{smallmatrix}]$. CMM used an initial estimate of R as $R_0 = I_{2\times 2}$, and the initial time instant for matrix estimation as $T/2$. VBM used the parameters for the inverse Gamma distribution as $\alpha = [0 ; 0]$ and $\beta = [1;1 ]$. The number of filtering iterations was selected as 2. The behavior of the precision factor was chosen as $\rho(n^t) = 1-min(50/n^t,0.1)$. IOCM used an initial condition for estimating $B$ as $B_0 = [0,0,0,0]$.


\bibliographystyle{IEEEtran}
\footnotesize
\bibliography{main}

\end{document}